\documentclass[journal]{IEEEtran}
\usepackage{cite}
\usepackage{amsmath,amssymb,amsfonts}
\usepackage{threeparttable}
\usepackage{enumerate}

\usepackage{algorithmic}
\usepackage{algorithm}
\usepackage{graphicx}
\usepackage{textcomp}
\usepackage{bm}
\usepackage{cases}
\usepackage{multirow}
\usepackage{longtable}
\usepackage{rotating}

\usepackage{multirow}
\usepackage{mathrsfs}
\usepackage{booktabs}
\usepackage[table]{xcolor}
\usepackage{threeparttable}
\usepackage{soul}
\soulregister\cite7
\soulregister\ref7
\ifCLASSINFOpdf
\else
\fi
\UseRawInputEncoding
\usepackage{lipsum}
\usepackage{fancyhdr}

\fancypagestyle{sec}{\lhead{\tmpx}}

\fancypagestyle{arXiv}
{
	\fancyhf{}
	
	\chead{\textcolor{gray}{This paper has been accepted for publication at the \\ IEEE Transactions on Computer-Aided Design of Integrated Circuits and Systems,  2022}}
	
	\cfoot{\vspace{-6mm} \fontsize{=6pt}{10pt}\selectfont  \textcolor{gray}{© 2022 IEEE. Personal use of this material is permitted. Permission from IEEE must be obtained for all other uses, in any current or future media, including reprinting/republishing this material for advertising or promotional purposes, creating new collective works, for resale or redistribution to servers or lists, or reuse of any copyrighted component of this work in other works}\\\fontsize{=10pt}{12pt}\selectfont\thepage}
	
}

\usepackage{tikz}
\newcommand*{\circled}[1]{\lower.7ex\hbox{\tikz\draw (0pt, 0pt)%
    circle (.5em) node {\makebox[1em][c]{\small #1}};}}

\begin{document}

\title{Skydiver: A Spiking Neural Network Accelerator Exploiting Spatio-Temporal Workload Balance
\thanks{
$^{\star}$ Qinyu Chen and Chang Gao are co-first authors. Qinyu Chen is the corresponding author.}
\thanks{
$^{1}$Qinyu Chen, Xinyuan Fang, Haitao Luan are with Institute of Photonic Chips, University of Shanghai for Science and Technology, Shanghai 200093, China,
University of Shanghai for Science and Technology, 200093 Shanghai, China.(Email:qinyu@usst.edu.cn)}
\thanks{
$^{2}$Chang Gao is with the Institute of Neuroinformatics, University of Z\"urich and ETH Z\"urich, 8057 Z\"urich, Switzerland}
}

 \author{\IEEEauthorblockN{Qinyu Chen$^{\star, 1}$,~\IEEEmembership{Member,~IEEE}, Chang Gao$^{\star, 2}$,~\IEEEmembership{Member,~IEEE},\\ Xinyuan Fang$^{1}$, Haitao Luan$^{1}$}\\

\texttt{}
 \vspace{-1.4cm}
}

\maketitle

\begin{abstract}
Spiking Neural Networks (SNNs) are developed as a promising alternative to Artificial Neural networks (ANNs) due to their more realistic brain-inspired computing models.
SNNs have sparse neuron firing over time, i.e., spatio-temporal sparsity; thus, they are useful to enable energy-efficient hardware inference.
However, exploiting spatio-temporal sparsity of SNNs in hardware leads to unpredictable and unbalanced workloads, degrading the energy efficiency.
In this work, we propose an FPGA-based convolutional SNN accelerator called \textbf{Skydiver} that exploits spatio-temporal workload balance.
We propose the Approximate Proportional Relation Construction (APRC) method that can predict the relative workload channel-wisely and a Channel-Balanced Workload Schedule (CBWS) method to increase the hardware workload balance ratio to over 90\%.
Skydiver was implemented on a Xilinx XC7Z045 FPGA and verified on image segmentation and MNIST classification tasks.
Results show improved throughput by 1.4$\times$ and 1.2$\times$ for the two tasks.
Skydiver achieved 22.6\,KFPS throughput, and 42.4 \textmu J/Image prediction energy on the classification task with 98.5\% accuracy. 

\end{abstract}


\IEEEpeerreviewmaketitle
\texttt{}
 \vspace{-1.3cm}

\section{Introduction}
\thispagestyle{arXiv}
\IEEEPARstart{O}{ver} the past decade, the revolution of Deep Neural Networks (DNNs) has led to an impressive performance on various challenging tasks.
However, such continuous-valued networks usually have tremendous parameters, leading to a large memory footprint and power budget when deployed on resource-constrained platforms.
To deal with this problem, many researchers designed DNN compression methods and efficient hardware architectures~\cite{9043731}.
Another promising approach is to use Spiking Neural Networks (SNNs).
Compared to non-spiking DNNs,
SNNs mimic the spiking behavior of biological neurons,
thus have the potential to achieve better performance with lower complexity.
The energy efficiency primarily benefits from:
(1) the sparsity brought by discrete spikes;
(2) replacing multiply-accumulate (MAC) operations by addition due to binary spike connections.
However, modern silicon implementations of SNNs have lagged behind those of DNNs, mainly featuring lower throughputs~\cite{2020iscaspinflow}.
Since SNNs run over multiple timesteps,
the sparsity exists spatially across neurons and temporally over timesteps, i.e., spatio-temporal sparsity.
However, exploiting spatio-temporal sparsity in hardware usually leads to an unpredictable and dynamic workload, affecting the hardware efficiency of running SNNs.

Research on improving SNN efficiency can be categorized into three categories.
The first category is to deploy SNNs on commercial architectures.
However, CPUs lack enough parallelism to achieve decent throughput.
GPUs have high efficiency when running tasks with highly parallel computation and memory access, but low efficiency when facing irregular computational flow and data access which is ubiquitous in SNNs.
The second approach is to design domain-specific hardware,
including large-scale systems (e.g. Loihi~\cite{2018Loihi}, TrueNorth~\cite{akopyan2015truenorth}) and low-power accelerators~\cite{neil2014minitaur,sen2017approximate,9491036,fang2020encoding,8954866,ju2020fpga}. Our work also falls into this category.
The third category is to explore emerging devices that are easier adaptable to the event-driven properties of SNNs~\cite{zhou2021large,wijesinghe2018all}.

\begin{figure}[t]
    \centering
    \includegraphics[width=0.5\textwidth]{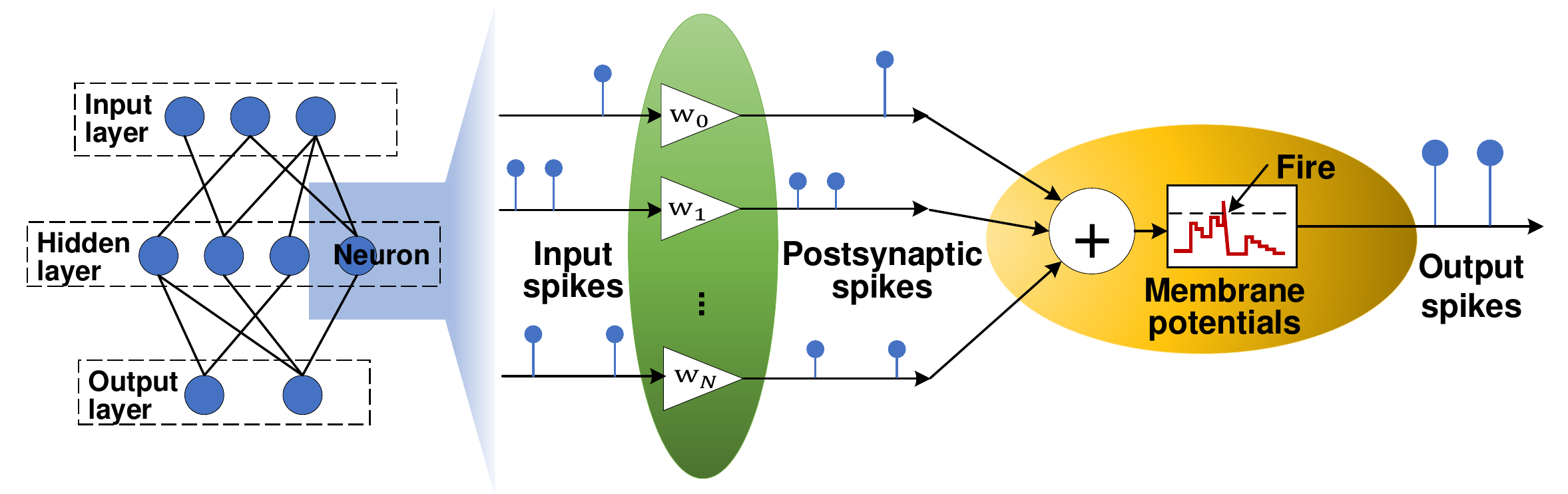}
    \caption{SNN topology and spiking neural dynamics
}
    \label{fig:snn-basics}
\end{figure}

\begin{figure*}[t]
    \centering
    \includegraphics[width=0.85\textwidth]{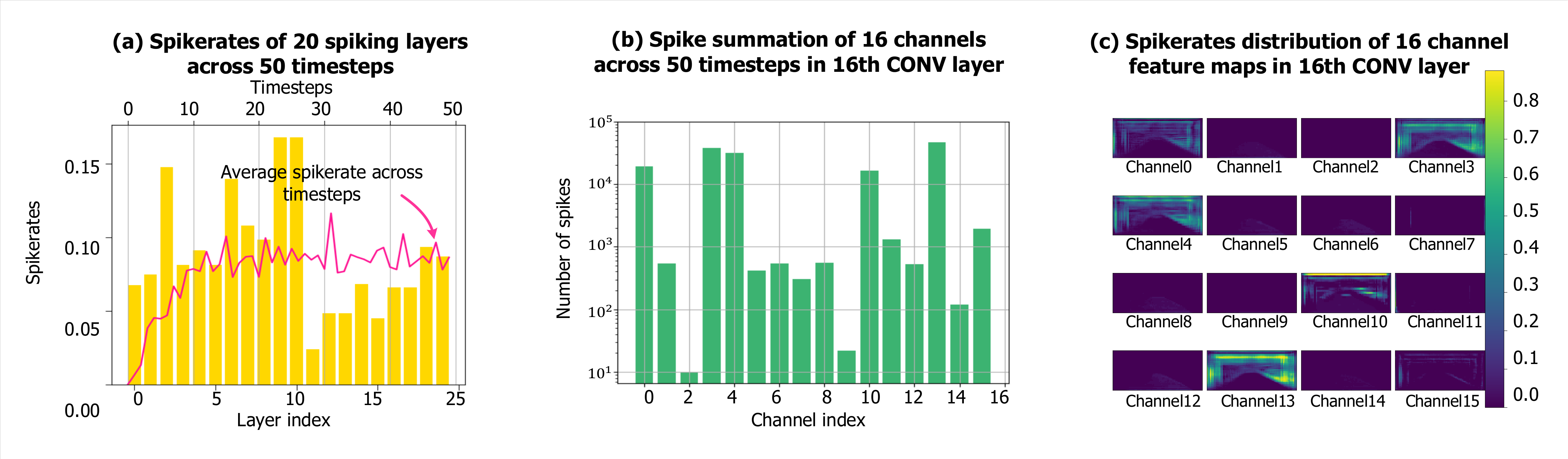}
    \caption{(a) Spikerates of each spiking layer, (b) spike summation of each output channel in the 16th layer, (c) spike rates distribution of each output channel in the 16th layer. The data are collected during segmenting a frame in a driving video, and the output is the segmentation mask for the road.
}
    \label{fig:motivation}
\end{figure*}

This paper proposes an SNN accelerator, Skydiver, that achieves high throughput by exploiting the spatio-temporal workload balance.
The main contributions are:
\begin{itemize}
\item We proposed an Approximate Proportional Relation Construction (APRC) method to predict the relative workload channel-wisely offline by modifying the network structure without any accuracy loss.
\item Based on the workload prediction, our Channel-Balanced Workload Schedule (CBWS) method can be easily applied to exploit spatio-temporal workload balance.
\item The proposed accelerator Skydiver is implemented on a Xilinx XC7Z045 FPGA and verified by image segmentation and classification tasks.
Results show that Skydiver achieved 98.5\% classification accuracy and 22.6K~FPS throughput on the MNIST dataset.
\end{itemize}
\texttt{}
 \vspace{-0.6cm}
\section{Motivation}
As shown in Fig.~\ref{fig:snn-basics}, SNNs are organized in cascaded layers and are executed over timesteps with inputs encoded in spike trains.
A spike generated by a neuron triggers the update of the membrane potential of each fan-out neuron.
The membrane potential $V^l_i(t)$ integrates the input current $z$ at each time step:
\begin{equation}
V^l_i(t) = V^l_i(t-1) + z^l_i(t)-V_{th}\Theta^l_i(t),
 \label{eq1}
\end{equation}
\noindent where $\Theta^l_i(t)$ is a spike, $V_{th}$ is the voltage threshold, and $z^l_i(t)$ is the input of neuron $i$ in layer $l$, which is given as:
\begin{equation}
z^l_i(t) = \sum_{j=1}^{M^{l-1}}W^l_{ij}\Theta^{l-1}_j(t) + b^l_i,
 \label{eq2}
\end{equation}
\noindent where $W^l_{ij}$ is the synaptic weight. 
A spike $\Theta$ is generated when $V^l_i(t)$ exceeds the threshold $V_{th}$ as
\begin{equation}
\Theta^l_i(t) = U(V^l_i(t) -V_{th}\Theta^l_i(t-1)),
 \label{eq3}
\end{equation}
\noindent where $U(x)$ denotes a unit step function.
The spikes propagate through the network until reaching the output.

SNNs have intrinsically event-driven workloads since the update of membrane potentials is triggered by spikes.
Fig.~\ref{fig:motivation}(a) shows that the spikerate varies, ranging substantially from 2\% to 18\% across the layers.
The average spikerate is less than 8\%, indicating a high level of spatio-temporal sparsity.
Thus, arithmetic operations can be saved if connections without a spike are skipped.
Moreover, the spikerate differs across timesteps, indicating that the proportion of active neurons varies over time.
The dynamic active connections between neurons introduce an irregular computation flow and unpredictable memory access patterns, leading to an unbalanced workload.
Fig.~\ref{fig:motivation}(b) and~\ref{fig:motivation}(c) showcase the unbalanced workload with respect to channels. They present the spike summation and distribution over 50 timesteps of 16 channels in a representative spiking layer, from which we find the significant imbalance by several orders of magnitude.
Next, we present how Skydiver balances the workload to achieve higher hardware efficiency.

\section{Architecture Design and Data Processing}
To exploit spatio-temporal workload balance, we first create an approximately proportional relationship between channel spikerates and filter magnitudes to predict the relative channel workload. Then, a channel-balanced workload schedule method is applied on Skydiver,
which can alleviate the imbalance without overheads.

\subsection{Skydiver: Architecture}
\begin{figure}[tbp]
    \centering
    \includegraphics[width=0.45\textwidth]{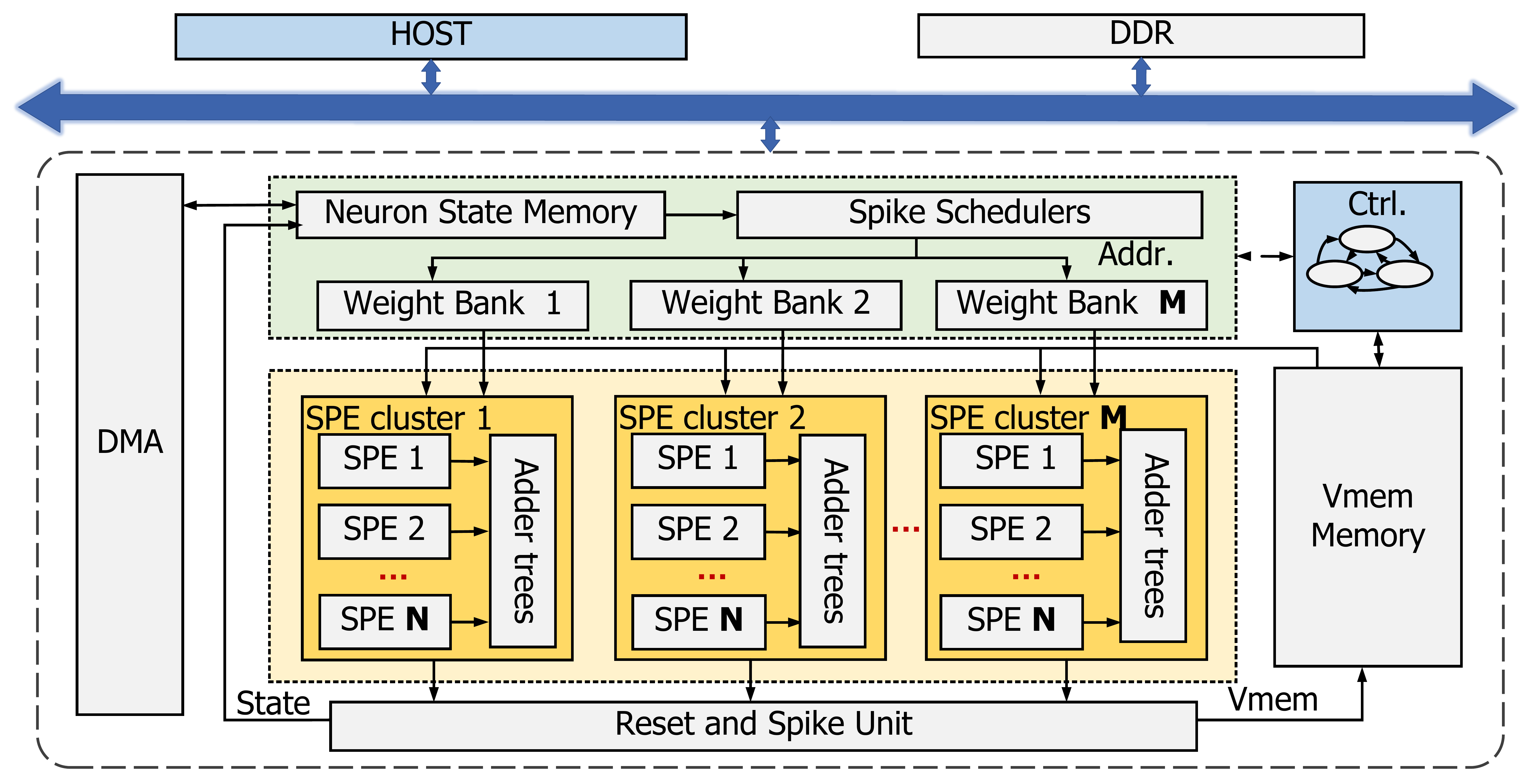}
    \caption{The proposed system architecture.
}
    \label{fig:block-diagram}
\end{figure}

The top-level architecture of the proposed Skydiver accelerator is depicted in Fig.~\ref{fig:block-diagram}.
Skydiver is composed of a controller, memory blocks, and processing elements.
The memory module contains a neuron state memory,
a membrane potential (VMEM) Memory,
and weight memory.
A Xilinx Direct Memory Access (DMA) IP block controlled by the host is used to manage the I/O communications between the accelerator and the host.
Input spike trains are streamed from DDR to the accelerator and buffered in the neuron state memory.
The spike scheduler is used to detect the neurons that fire and generate the memory address of the corresponding weights, of which the details can be found in our previous work~\cite{9491036}.
The processing elements comprise several filter-based Spiking Processing Elements (SPE) clusters.
Each SPE cluster is connected to a corresponding weight bank and contains multiple channel-based SPEs,
the finest grain of workload balance.
The controller manages the overall execution,
which updates the state of the accelerator and decodes the information fetched from the host.

\subsection{Approximate Proportional Relation Construction (APRC) between Channel Spikerates and Filter Magnitude}

\begin{figure}[tbp]
    \centering
    \includegraphics[width=0.43\textwidth]{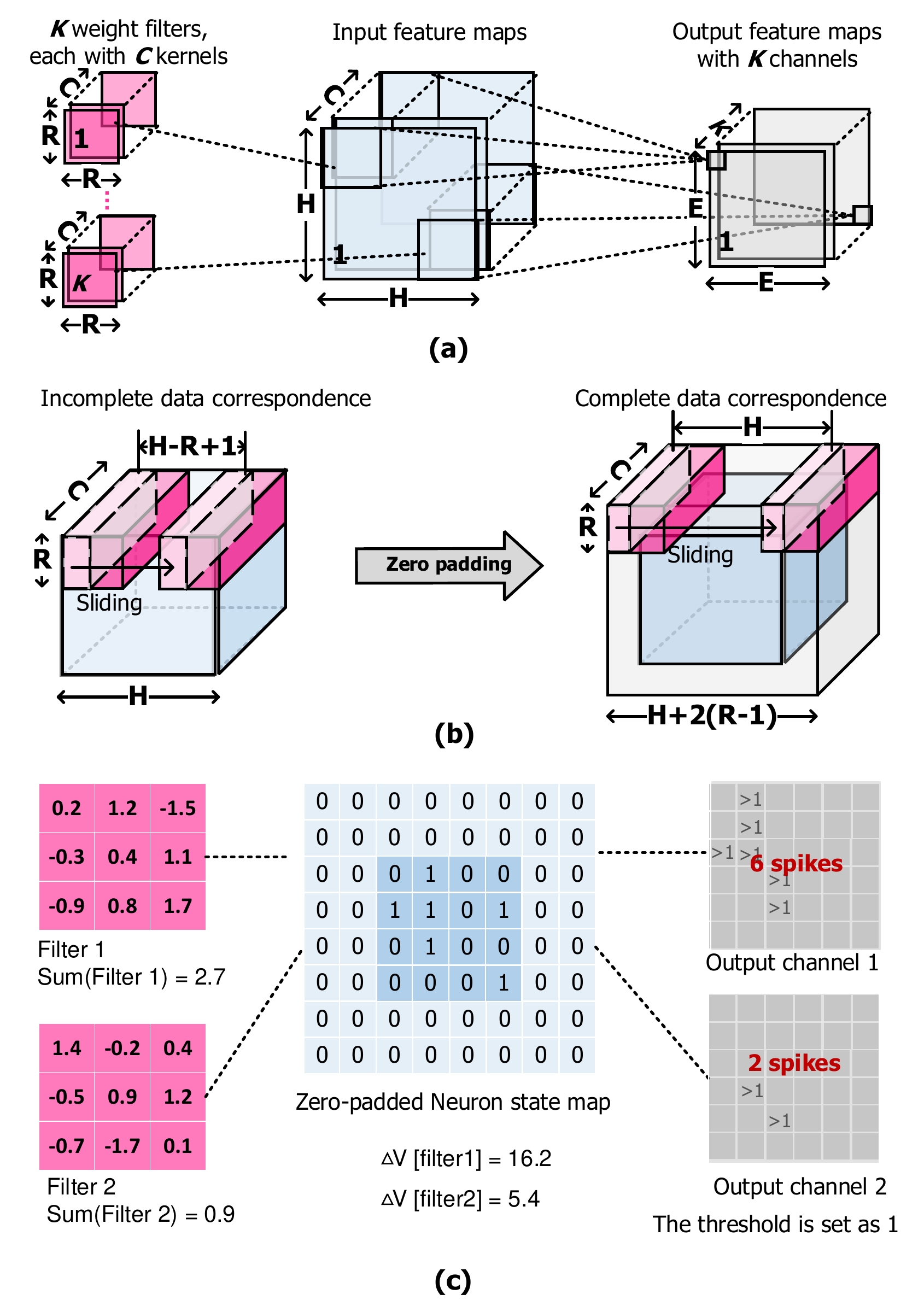}
    \caption{(a) The topology of convolutions, (b) modified convolutions, (c) an example of proportional relation construction between channel spikerates and filter magnitude based on modified convolutions.
}
    \label{fig:linear-link}
\end{figure}

Fig.~\ref{fig:linear-link}(a) outlines how the output feature maps are formed by filters and input feature maps in a convolutional layer.
The core operation is a 3-dimensional sliding window convolution of a $R$$\times$$R$$\times$$C$ element $filter$ over a $H$$\times$$H$$\times$$C$ element $input$ $channels$.
Multiple ($M$) filters can be applied to the same input feature maps to generate $M$ $output$ $channels$,
each output channel is associated with a filter.

According to Eq.(1)-Eq.(3), the larger the filter magnitude (i.e., the summation of all the elements within a filter) is, the higher membrane potential the neuron has accumulated.
However, shown in Fig.~\ref{fig:linear-link}(b), the exact proportionality could not be met by the normal convolution due to the incomplete data correspondence and the uneven data distribution.
To solve this problem, $(R-1)$ zeros need to be padded around every channel,
and the stride $S$ should be set as 1.
This ensures that each element within a filter can be applied to all input neuron state map elements.

The updated membrane potential $\Delta V$ of timestep $t$ at location $(x,y)$ in output feature map $n$ is given by:
\begin{equation}
\Delta V_{n}[t] = \sum_{i=0}^{C-1}\sum_{j=0}^{R-1}\sum_{k=0}^{R-1} (w_{n}[i][j][k] \times in[t][i][x+j][y+k]),
 \label{eq1}
\end{equation}
The summation of the updated membrane potential across the output channel $n$ can be expressed as
\begin{equation}
\begin{split}
\sum_{x=0}^{E-1}\sum_{y=0}^{E-1}\Delta V_{n}[t] = (\sum_{i=0}^{C-1}\sum_{j=0}^{R-1}\sum_{k=0}^{R-1} w_{n}[i][j][k]) \\
 \times (\sum_{i=0}^{C-1}\sum_{x=0}^{H-1}\sum_{y=0}^{H-1}in[t][i][x+j][y+k]),
 \label{eq1}
 \end{split}
\end{equation}
\noindent from which we can observe the exactly proportional relationship between the magnitude of the filter $n$ and the summation of the updated membrane potential across the corresponding output channel.
Considering that the distribution and the reset mechanism in SNNs may have an impact on the effect of exact proportionality,
APRC can make the relationship between the channel spikerates and filter magnitudes approximately proportional.

For better illustration, we take an example (Fig.~\ref{fig:linear-link}(c)) to explain how the proportional relation between channel spikerates and filter magnitude are constructed based on modified convolutions.
Assume two $3\times3$ filters with different magnitudes (2.7 and 0.9; the ratio is 3) and an $8\times8$ input neuron state map padded by zeros. After convolutions, two output channels are generated.
The summations of the updated membrane potential across the two output channels are 16.2 and 5.4, respectively, showing the same ratio as that of filter magnitudes.
A spike is generated once the membrane potential exceeds the pre-defined threshold.
6 and 2 spikes are generated in the two output channels, respectively. It shows that the filter magnitudes have a proportional relationship with output channel spikerates.

\subsection{Channel-Balanced Workload Schedule (CBWS) mechanism}
\begin{figure}[tbp]
    \centering
    \includegraphics[width=0.45\textwidth]{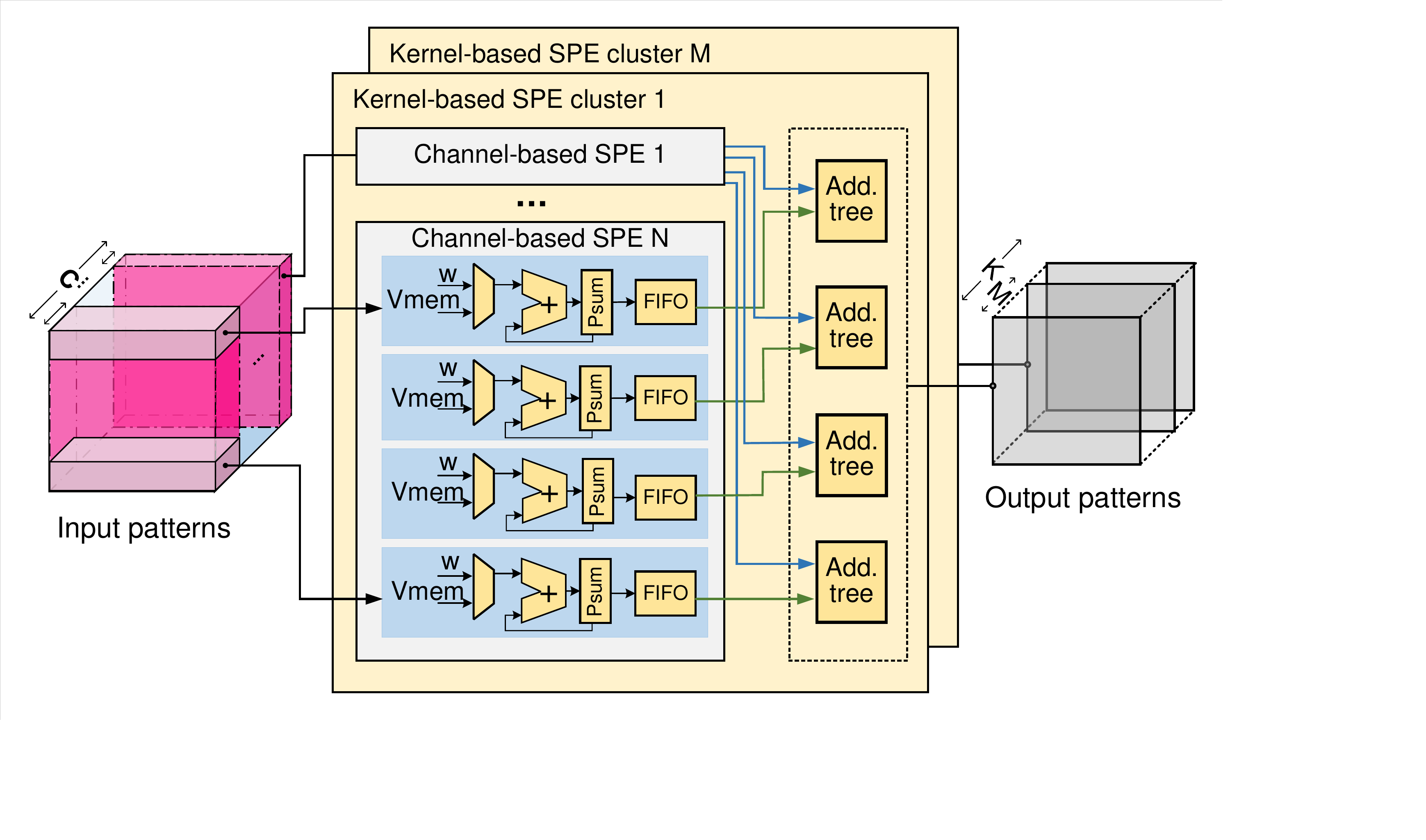}
    \caption{Structure of the filter-based spiking processing unit.
}
    \label{fig:kspu}
\end{figure}

\begin{algorithm}[thpb]
\caption{Channel-balanced workload schedule mechanism}
\label{alg:2}
\begin{algorithmic}[1]
\REQUIRE ~~\\ 
$W$, the parameters in the network;
$K$, the number of filters in a certain layer;
$M$, the number of SPE clusters;
$N$, the number of SPEs in a cluster.
\ENSURE ~~\\ 
Balanced task allocation; \\
\STATE Sum up the parameters of each kernel, $s = \sum W$.
\STATE Build a list $C$ containing all the $s$, and sort the elements in descending order.
\STATE $//$ Build a new list $C_{new}$ by resorting list $C$ piecewisely.
\FOR{$i=0$; $i \le K/N-1$; $i++$ }
\IF {$mod(i,2)$}
\STATE $C_{new}.append(C[Ni:Ni+N-1])$.
\ELSE
\STATE $C_{new}.append(sort(C[Ni:Ni+N-1], 'descend')$.
\ENDIF
\ENDFOR
\STATE $//$ Split the list $C$ into $N$ sublists, and initialize them with roughly equal summations.
\FOR{$i=0$; $i \le K/N-1$; $i++$ }
\FOR{$j=0$; $j \le N-1$; $j++$ }
\STATE Add the element $C_{new}[Ni+j]$ to sublist $L_{j}$;
\ENDFOR
\ENDFOR
\STATE  $//$ Finetune the elements in sublists within $T$ iterations.
\FOR{$t=0$; $t \le T$; $t++$ }
\STATE $sum_j = \sum L_j$.
\STATE $diff(t) = max(sum_j) - min(sum_j)$.
\STATE Obtain the sublist $L_{max}$ with $max(sum_j)$, and the sublist $L_{min}$ with $min(sum_j)$.
\IF {$diff(t)/2 > min(L_{max})$}
\STATE move the element $min(L_{max})$ from $L_{max}$ to $L_{min}$.
\STATE Update $L_i$.
\ELSE
\STATE  BreakTimeLoop()
\ENDIF
\ENDFOR

\end{algorithmic}
\end{algorithm}

SPE clusters are provided with different filters, and each cluster can calculate the membrane potential of a specific output channel independently.
As shown in Fig.~\ref{fig:kspu}, a cluster consists of several channel-based SPEs and adder trees.
Each channel-based SPE receives a subset of $kernels$ within a filter
and produces partial sums of membrane potentials.
Within a channel-based SPE, the workload is further partitioned into four streams, which calculate equal rows of elements in the output.
Each adder tree collects the partial sums from the corresponding stream of all channel-based SPEs.

Since the sparsity of input channels are quite different, the SPE which computes channels with the most zero inputs first completes the task,
whereas the SPE computing channels with the most nonzero inputs become the bottleneck of hardware throughput.
To deal with this unbalanced event-driven workloads,
the Channel-Balanced Workload Schedule (CBWS) method (seen in Algorithm~\ref{alg:2}) is proposed,
in which the input channels will be partitioned into $N$ groups with almost equal workloads and processed by $N$ SPEs.
First, build a list containing filter magnitudes;
Second, resort the list piecewisely by making each two adjacent data fields have opposite orders;
Third, split the list into $N$ sublists, and initialize them with roughly equal element summations by gathering the elements with the same index in each data field together; Finally, fine-tune the elements in each sublist.

\section{Experimental Results}
Skydiver was evaluated on an image classification task and a segmentation task.
As for the classification model, we adopt a spiking network with a 28x28-16c-32c-8c-10 structure on the MNIST dataset.
As for the spiking segmentation model,
we use the structure with 160x80x3-8C3-16C3-32C3-32C3-16C3-1C3-160x80x1 from MLND-Capstone project\footnote{https://github.com/mvirgo/MLND-Capstone/},
which has 189.5K parameters.

\begin{figure}[t]
    \centering
    \includegraphics[width=0.4\textwidth]{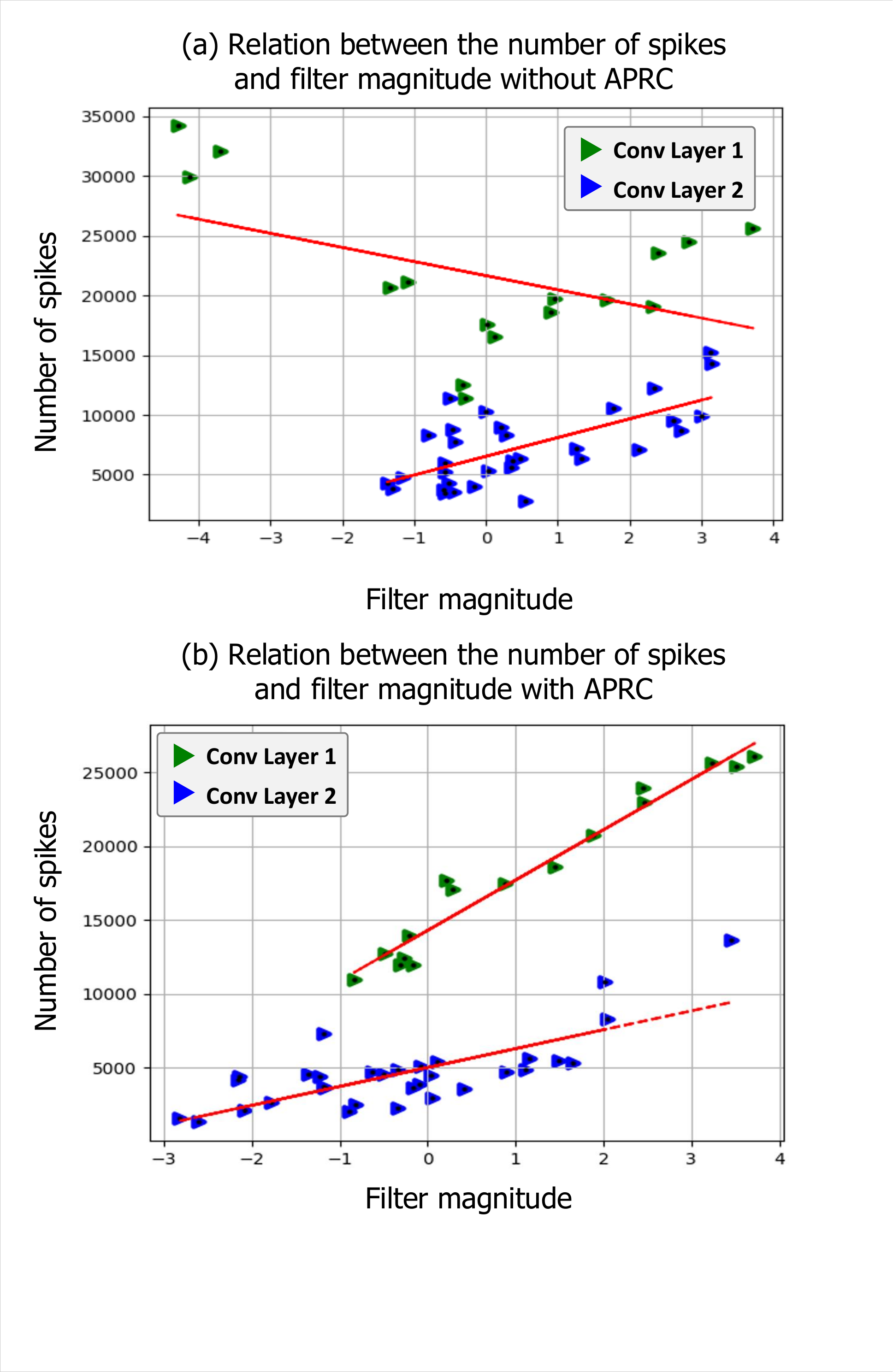}
    \caption{The relation between the number of spikes in an output channel and the filter magnitude of typical CONV layers in the classification network.
}
    \label{fig:appr-linear}
\end{figure}

The proposed APRC strategy is supposed to construct an approximate proportional relation between the number of spikes in an output channel and the corresponding filter magnitude.
Fig.~\ref{fig:appr-linear}(a) shows a relatively irregular relation between the number of spikes and filter magnitude when APRC was not applied.
Fig.~\ref{fig:appr-linear}(b) shows that APRC constructed an approximately proportional relation in the convolutional SNN layers of the classification network.
\begin{figure}[t]
    \centering
    \includegraphics[width=0.38\textwidth]{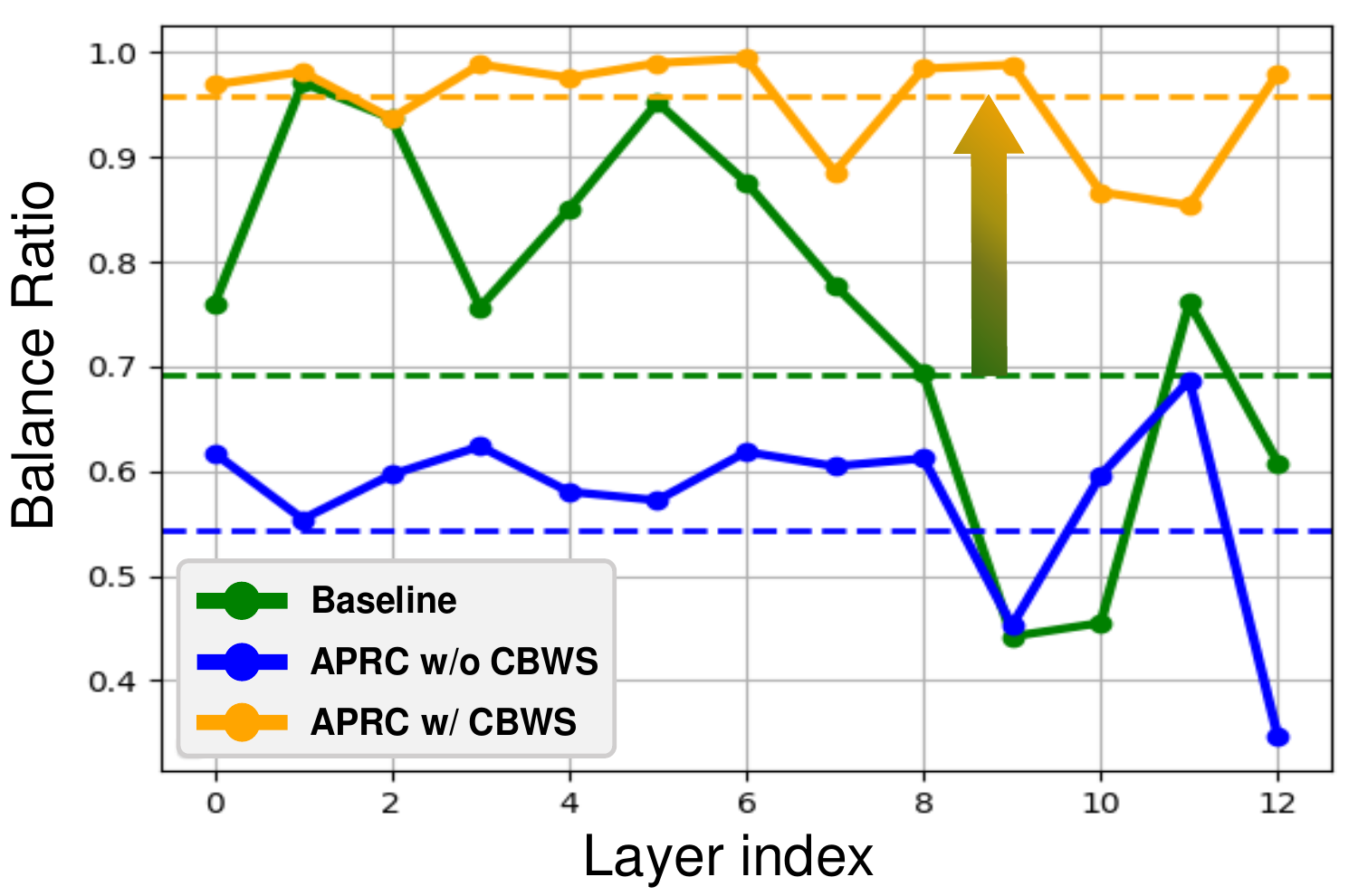}
    \caption{Balance ratio of the Skydiver accelerator versus layer index in a segmentation network, which is evaluated on MLND-Capstone dataset. The dotted lines denote the average level across layers.
}
    \label{fig:br}
\end{figure}
\begin{table*}[t]\renewcommand\arraystretch{1.1}
	\centering
    \begin{threeparttable}[b]
	\caption{Comparison with Previous Works}
\label{tab:final}
\setlength{\tabcolsep}{3mm}{
\begin{tabular}{c|c|c|c|c|c}
\hline
\hline
Metrics  &  TCAS-I'21\cite{li2021fast} & ICCAD'20\cite{fang2020encoding}   & ASSCC'19\cite{zhang2019asynchronous}   & Neural Comp.'20\cite{ju2020fpga}&This work \\
\hline
Platform    &  VC707   &XCZU9EG       &XC7VX690T    &Zynq ZCU102    & XC7Z045\\
\hline
Network      & MLP & MLP/CNN\footnotemark[1]& MLP\footnotemark[2] &CNN\footnotemark[3] &CNN/CNN\\
\hline
Task        & image classifi. & image classifi.& image classifi.& image classifi. & image classifi./video seg. \\
\hline
Freq. (MHz)    &$100$  &$125$            &-     &$100$  &$200$\\
\hline
on-chip Power (W)   &$1.6$ &$4.5$  & $0.7$ & $4.6$ & $0.96$\\
\hline
Predition Energy (mJ/frame)  &5.04 &2.34/33.84       & 0.77  & 30 &9.12@seg./0.04@classif. \\
\hline
KFPS & 0.32& 1.92/0.13   & 0.91   &0.16 &1.04  \\
\hline
Throughput (GSOp/s) &$-$ &$-$  & $0.73$ &$-$& 0.11@seg./ 22.6@classif. \\
\hline
Efficiency (GSOp/s/W) &$-$ &$-$  & $0.95$ &$-$ & $19.3$\\
\hline
\hline
\end{tabular}}
\begin{tablenotes}
       \footnotesize
       \item[1] Classification network with 784-500-500-10 and 28x28-32C3-P2-32C3-P2-256-10 for MNIST.
       \item[2] Classification network with 784-512-384-10 for MNIST.
       \item[3] Classification network with 28x28-64C5-2S-64C5-2S-128F-10 for MNIST.
\end{tablenotes}
\end{threeparttable}
\end{table*}
\begin{table}[t]\renewcommand\arraystretch{1.3}
	\centering
    \begin{threeparttable}[b]
	\caption{XC7Z045 FPGA resource utilization of Skydiver}
\label{tab:resource}
\setlength{\tabcolsep}{4mm}{
\begin{tabular}{ccccc}
\hline
Metrics& LUT & FF &DSP & BRAM  \\
\hline
\hline
Avaliable &218600 &437200        &900     &545 \\
\hline
Used    & 45986 &20544            & 0      &262    \\
\hline
Percentage    &21.04\%  &4.70\%   &0\% &48.07\%  \\
\hline
\end{tabular}}
\end{threeparttable}
\end{table}
Unbalanced workloads between SPEs lead to deteriorated actual throughput. Thus, the most important feature of Skydiver is the ability to efficiently handle the dynamic workload of SNNs, which is quantified by the \textit{balance ratio} defined in~\cite{gao2021spartus}.
Fig.~\ref{fig:br} shows the balance ratio across layers of the segmentation network.
It was observed that the performance of the CBWS mechanism is directly affected by the existence of the APRC strategy.
When CBWS has applied alone,
this design only achieved a 54.37\% balance ratio.
After combining with the APRC mechanism,
the ratio was improved to 95.69\%.
Besides, without the proposed CBWS and APRC strategies,
the accelerator achieved a 69.19\% ratio,
which proves the importance of APRC in predicting workloads in advance.
The balance ratio is improved for the classification network from 79.63\% to 94.14\% using both APRC and CBWS.
The higher balance ratios result in 1.4$\times$ and 1.2$\times$ actual throughput increase in the segmentation and classification tasks, respectively.

The proposed Skydiver accelerator is synthesized and implemented on an XC7Z045 FPGA at 200~MHz.
The host manages the input and output data transfer of the programmable logic.
The resource utilization of the programmable logic, which reflects the physical area of Skydiver, is shown in Table~\ref{tab:resource}.
Table~\ref{tab:final} presents the results of this work and prior state-of-the-art SNN processors.
Skydiver achieved $110$~FPS throughput and 0.91 mJ/image prediction energy when processing the segmentation network,
and $22.6$K~FPS throughput, 42.4~$\mu$J/image prediction energy when processing the classification network.
Compared to prior SNN processors, Skydiver achieved competitive prediction throughput.
Prior DNN accelerators such as Sparten~\cite{gondimalla2019sparten} groups filters by density to schedule the workloads, however it can not solve the workload imbalance caused by dynamic neuron state sparsity in SNNs.

\section{Conclusion}
In this work, we introduced an energy-efficient convolutional SNN accelerator called Skydiver.
We showed that a significant workload imbalance exists in the channels of the feature maps.
To exploit spatio-temporal workload balance, the APRC method was proposed to predict the relative workload channel-wisely.
Based on the prediction, the workload balancing method CBWS was proposed to improve the hardware efficiency and throughput gain.
Skydiver was evaluated on an XC7Z045 FPGA, and the results show that it can achieve competitive energy efficiency and throughput.

\bibliographystyle{IEEEtran}

\bibliography{IEEEabrv,NOAH}

\end{document}